\documentclass[10pt,twocolumn,twoside]{IEEEtran} 

\IEEEoverridecommandlockouts                  
\usepackage{amssymb,amsmath,amsthm, amsfonts} 
\usepackage{amssymb}  
\usepackage{graphicx,amsmath,amssymb}
\usepackage{cite}
\usepackage{subfigure}
\usepackage{comment,color}

\newcommand{\R}{\ensuremath{{\mathbb R}}}

\newcommand{\Z}{{\mathbb Z}}

\newtheorem{thm1}{\bf Theorem}

\newtheorem{lem1}{\bf Lemma}
\newtheorem{assmpt1}{\bf Assumption}
\newtheorem{defn1}{\bf Definition}
\newtheorem{rem1}{\bf Remark}

\newtheorem{cor1}{\bf Corollary}

\newenvironment{asm}{\begin{assmpt1}}{\hfill\end{assmpt1}}
\newenvironment{rem}{\begin{rem1}}{\hfill\end{rem1}}
\newenvironment{lem}{\begin{lem1}}{\hfill\end{lem1}}
\newenvironment{thm}{\begin{thm1}}{\hfill\end{thm1}}

\usepackage{algorithm,algorithmic}

\title{\LARGE \bf
	Fully Decentralized Design of Initialization-free\\ Distributed Network Size Estimation
}

\author{Donggil Lee, Taekyoo Kim, Seungjoon Lee, and Hyungbo Shim
	\thanks{This work was supported by the National Research Foundation of Korea(NRF) grant funded by the Korea government(Ministry of Science and ICT) (No. RS-2022-00165417).}
	\thanks{Donggil Lee is with Center for Intelligent and Interactive Robotics, Korea Institute of Science and Technology, Seoul 02792, Korea {\tt\small(dglee@kist.re.kr)}. Taekyoo Kim is with the Train Control and Communication Team, Korea Railroad Research Institute, Uiwang 16105, Korea {\tt\small  (tkkim13@krri.re.kr)}. Seungjoon Lee is with NAVER LABS, Seongnam 13161, Korea {\tt\small  (seungjoon.lee@naverlabs.com)}. Hyungbo Shim is with ASRI, Department of Electrical and Computer Engineering, Seoul National University, Seoul 08826, Korea {\tt\small  (hshim@snu.ac.kr)}.}
}

\begin{document}
	
	\maketitle
	\thispagestyle{empty}
	\pagestyle{empty}
	
	\begin{abstract}
		In this paper, we propose a distributed scheme for estimating the network size, which refers to the total number of agents in a network. By leveraging a synchronization technique for multi-agent systems, we devise an agent dynamics that ensures convergence to an equilibrium point located near the network size regardless of its initial condition. Our approach is based on an assumption that each agent has a unique identifier, and an estimation algorithm for obtaining the largest identifier value. By adopting this approach, we successfully implement the agent dynamics in a fully decentralized manner, ensuring accurate network size estimation even when some agents join or leave the network. 
	\end{abstract}


	\section{Introduction}\label{sec:Intro}
	Recent advancements in network communication technologies have prompted extensive research efforts  on distributed algorithms. These studies involve the cooperation of multiple agents striving to achieve a common goal under a localized communication structure, wherein each agent can exchange information with its neighboring agents. Such structural setup provides significant benefits, including scalability, robustness against failures, and structural flexibility. Extensive research and application of distributed algorithms can be found across various problem domains, such as distributed optimization \cite{nedic09TAC, nedic10TAC}, distributed state estimation \cite{kim16CDC, wang17TAC}, and formation control \cite{anderson08CSM, oh15Aut}.
	
	The term ``distributed'' naturally leads to the expectation that the distributed algorithms utilize only local knowledge for both their design and operation. However, many distributed algorithms often require global information, such as the {\it network size} or the {\it algebraic connectivity}\footnote{The algebraic connectivity refers to the second-smallest eigenvalue of the Laplacian matrix for the underlying graph (also known as the Fiedler eigenvalue).}. For instance, distributed estimation \cite{kim16CDC2}, distributed control \cite{kim2023decentralized}, and leader-following consensus \cite{ren2021optimal} all utilize the exact value of the network size for their filter design. Similarly, achieving synchronization in systems that range from simple, linear, and homogeneous \cite{tuna08arXiv,seo09Aut} to complex, nonlinear, and heterogeneous \cite{kim15TAC,panteley17TAC,yun19Aut} necessitates knowledge of a lower bound for the algebraic connectivity to determine suitable parameters.
	
	Under mild assumptions on the communication network (see, for example, Assumption \ref{asm:graph}), one can obtain a lower bound\footnote{Refer to \eqref{eq:lboundOflambda2} for the explicit form, where $\lambda_2$ represents the algebraic connectivity, and $N$ denotes the network size.} for the algebraic connectivity using the network size. Therefore, distributed algorithms that estimate the network size can be an essential building block for the aforementioned algorithms, enabling their design with local information only.

	The purpose of this paper is to propose a distributed scheme for estimating network size with two key features:
	\begin{itemize}
		\item[(F1)] (Decentralized design) Each agent {\it self-organizes} its own algorithm using its local knowledge, without the need for global information or an external coordinator. 
		\item[(F2)] (Initialization-free) The algorithm does not rely on any specific initial conditions.
	\end{itemize}
	
	Attaining both features may be challenging, but they are aimed at realizing ``plug and play'' operation; that is, any agents can be effortlessly integrated into (or separated from) the network by simply plugging in (or unplugging) them. With the features (F1) and (F2) in place, new agents can participate in the network without any global information, and the remaining agents do not need to go through a re-initialization process.  Furthermore, even when some agents become disconnected from the network due to equipment problems, the algorithm continues to function robustly without requiring any special action. 
	
	As a design tool to achieve the goal, we employ the {\it blended dynamics} approach \cite{kim15TAC}. Consider a system of heterogeneous dynamics involving $N$ agents described by
	\begin{align}
		\dot {\hat x}_i = f_i(\hat x_i)+k\sum_{j\in\mathcal N_i}(\hat x_j-\hat x_i),~ i\in\mathcal N:=\{1,\cdots,N\}\label{eq:exampleForBlended}
	\end{align}
	where $\hat x_i\in\R$ is the state, $f_i:\R\rightarrow\R$ is the individual vector field, $\mathcal N_i$ denotes the index set of neighboring agents that can give information to agent $i$, and $k>0$ is the coupling gain. Under an undirected and connected communication graph, it has been shown that if the coupling gain $k$ is sufficiently large, then each state $\hat x_i(t)$ approximates the solution $s(t)\in\R$ of the blended dynamics given by
	\begin{align}
		\dot s = \frac{1}{N}\sum_{i=1}^N f_i(s)\label{eq:generalBlendedDynamics}
	\end{align}
	as long as the blended dynamics is contractive or incrementally stable.
	
	Inspired by this theory, one can design the individual vector fields $f_i$ as follows:
	\begin{align}\label{eq:blendedDynamics_N}
		f_i(\hat x_i)=\begin{cases}1-\hat x_i,\quad &\text{if }i=1,\\1,\quad &\text{otherwise}.\end{cases}
	\end{align}
	This choice yields the blended dynamics as
	\begin{align}
		\dot s = \frac1N(N-s).\label{eq:blendedDynamics}
	\end{align}
	Thus, the solution $s(t)$ converges to $N$ from any initial condition $s(0)$, and this suggests that each state $\hat x_i(t)$ will approach close to the network size $N$ with a sufficiently large coupling gain $k$. This idea was first adopted in a conference version of this paper \cite{lee18CDC} and successfully attains the feature (F2). However, it fails to achieve feature (F1) due to two reasons. First, it requires a prior election of a special agent (say, a leader) labeled as $i=1$, whose dynamics differs from others as described in \eqref{eq:blendedDynamics_N}. Therefore, the election of the leader was driven by an external coordinator rather than individual agents. Second, accurate estimation of the network size is guaranteed only when the coupling gain $k$ is greater than or equal to $N^3$. Hence, all agents must have knowledge of a shared upper bound $\bar N$ of network size $N$ in order to determine a worse case estimate of the coupling gain $k$ (e.g., $k = \bar N^3$).
	
	Unfortunately, it has been proven that the distributed election of a single leader is unsolvable in general, due to the inherent symmetry of communication networks, as detailed in \cite[Theorem 3.1.1]{santoro2006design}. To overcome this symmetry, we assign each agent a unique identifier, specifically as a positive integer value, which is reasonable since most communication devices have their own identifiers, such as MAC addresses. 
	
	We then focus on estimating the largest identifier. Once all agents have obtained this value, a leader can be chosen in a distributed fashion. That is, an agent declares itself as the leader if its identifier equals the largest identifier. The uniqueness of the identifiers ensures that only one leader emerges,  effectively resolving the first issue. Furthermore, the largest identifier serves as an upper bound for the network size, since there are at least $N-1$ positive integers smaller than it (e.g., all the identifiers excluding the largest identifier). Leveraging this fact, all agents choose their coupling gain as the cube of the largest identifier, addressing the second issue.
	
	Motivated by the above observations, we propose a distributed algorithm consisting of two components: one for estimating the largest identifier and another for estimating the network size using the outcome of the first part. Through this approach, all agents obtain the network size within a finite time, while attaining both key features (F1) and (F2).

	\subsection{Literature Survey}
	Among various methods for the network size estimation, probabilistic approaches have been in the spotlight. One area of research focuses on the {\it random walk strategy} \cite{gkantsidis2006random,massoulie2006peer} where a {\it token} is generated from one agent and exchanged in a random way through the network. Here, the network size is estimated based on the time elapsed from sending the token to arrival. Another research area involves the {\it capture-recapture strategy} \cite{petrovic2009new,peng2009estimation}, where one leader agent propagates a certain number of {\it seeds} through the network, and the network size is estimated by checking the ratio of agents having the seed. Additionally, some research relies on the order statistics of random numbers \cite{baquero2011extrema,varagnolo2013distributed,deplano2021dynamic}. In this approach, each agent generates an independent and identically distributed random number and computes the largest value of the random numbers using the maximum-consensus technique. Then, the network size is derived from the statistical relationship between the largest value and the network size. 
	
	However, all the aforementioned approaches suffer from scalability issues due to the stochastic nature of the estimation results. More specifically, as the network size increases, the variance of the estimated results also rises. To reduce such variance, one can conduct the above algorithms multiple times in parallel and then use the average of their estimates, but this would require an excessive amount of communication between the agents.	 
	
	On the contrary, deterministic approaches for estimating the network size have also been proposed. One strand of research relies on specific initialization process. For instance, the methods presented in papers \cite{shames2012distributed, kenyeres2019distributed, kenyeres2021performance} employ average-consensus techniques to compute the inverse of the network size by setting one leader agent's initial state to $1$ and the rest of the agents' states to $0$. While these methods are simple to execute, a cautious re-initialization of the states should be performed after each network switch.
	
	To eliminate such reliance on the initialization conditions, the paper \cite{lee18CDC} presents a distributed algorithm based on the blended dynamics theory. In this approach, equilibrium points of all agent dynamics are located close to the network size, independent of the initial conditions. This is achieved by embedding information regarding the network size in each agent's dynamics, rather than relying on initial conditions. The authors of the paper \cite{tran2022distributed} further extend this approach by employing the proportional-integral type coupling, which enables all the parameters to be designed with local information only. 	
	Despite such progresses, a challenging task remains to achieve the plug-and-play capability. In order to design all agents dynamics so that their blended dynamics is given as \eqref{eq:blendedDynamics}, it is necessary to elect a single leader with an agent dynamics that differs from those of other agents. As a result, if the leader agent leaves the network, the algorithms in \cite{lee18CDC,tran2022distributed} will not work unless a new leader is elected. Our contribution is to enable even this process to be conducted in a distributed manner. 

	\subsection{Notation and Preliminaries}
	We use the symbol $\delta_{ij}$ to represent the Kronecker delta, which equals 1 if $i=j$, and 0 if $i\neq j$. The vector $1_N\in\R^N$ consists of all ones, and the matrix $I_N\in\R^{N\times N}$ is the identity matrix. The Euclidean norm of a vector $x$ is denoted as $|x|$, and the induced  2-norm of a matrix $B$ is denoted as $\|B\|$. The cardinality of a finite set $C$ is represented by $|C|$. If a symmetric matrix $D$ is positive semi-definite, we write $D\succeq 0$. The set of all positive integers is represented by $\mathbb Z_+$. 
	
	Consider a communication topology of networked agents that can be represented by an unweighted graph $\mathcal G:=\{\mathcal N,\mathcal E\}$, where $\mathcal N:=\{1,\cdots,N\}$ is a finite set of the agents indices, and $\mathcal E\subset \mathcal N\times \mathcal N$ is a set of edges. An edge $(j,i)\in\mathcal E$ indicates that agent $j$ can give information to agent $i$. A set of neighboring agents that can give information to agent $i$ is denoted as $\mathcal N_i=\{j\in\mathcal N\,:\,(j,i)\in\mathcal E\}$. If $(j,i)\in\mathcal E$ implies $(i,j)\in\mathcal E$ for any $i,j\in\mathcal N$, then $\mathcal G$ is said to be undirected. A path from $j$ to $i$ is a sequence of indices $(i_0,\cdots,i_m)$ such that $i_0=j$, $i_m=i$, and $(i_k,i_{k+1})\in\mathcal E$ for all $k=0,\cdots,m-1$ with every $i_k$ being different. If there is a path from $j$ to $i$ for any two distinct indices $i,j\in\mathcal N$, then $\mathcal G$ is said to be connected. The {\it Laplacian matrix} of $\mathcal G$ is denoted as $\mathcal L=[l_{ij}]\in\R^{N\times N}$, where $l_{ij}$ is $|\mathcal N_i|$ if $i=j$, $-1$ if $j\in\mathcal N_i$, and 0 otherwise.
	
	Now, we provide useful relations for the matrix $\mathcal L$ when the graph $\mathcal G$ is undirected and connected. First, all the eigenvalues of  $\mathcal L$ are non-negative real numbers, and there is only one zero eigenvalue \cite{olfati2007consensus}, i.e., the eigenvalues can be sorted as $0=\lambda_1<  \lambda_2\leq \cdots\leq \lambda_N$, without loss of generality. Thus, Schur's lemma ensures that there exists a matrix $R\in\R^{N\times (N-1)}$ satisfying the followings:
	\begin{align}
		1_N^TR=0,\quad R^TR=I_{N-1},\quad R^T\mathcal LR = \Lambda\label{eq:aboutLaplacian}
	\end{align}
	where $\Lambda:=\text{diag}_{i=2}^N (\lambda_i)\in\R^{(N-1)\times(N-1)}$. Second, it follows from  \cite[Theorem 1]{anderson85LMA} that the matrix $\mathcal L$ satisfies
	\begin{align}\label{eq:uboundOfL}
		\|\mathcal L\|\leq N.
	\end{align}
	Third, a lower bound for $\lambda_2$ can be obtained from \cite[Theorem 4.2]{mohar91Springer}, as follows:
	\begin{align}\label{eq:lboundOflambda2}
		\lambda_2> \frac{4}{N^2}.
	\end{align}
	
	\subsection{Organization}
	The remainder of the paper is organized as follows. In Section \ref{sec:proposedAlgorithm}, details of the proposed algorithm are presented. In particular, we prove the exponential convergence of the proposed algorithm. Section \ref{sec:sim} shows the simulation results of the proposed scheme, and Section \ref{sec:conclusion} concludes this paper.
	
	\section{Distributed Algorithm for \\Network Size Estimation}\label{sec:proposedAlgorithm}	
	This section addresses the problem of network size estimation for interconnected agents. Consider a communication topology of the agents represented by an unweighted graph $\mathcal G=\{\mathcal N,\mathcal E\}$, where $\mathcal N=\{1,\cdots, N\}$ is a set of the agents indices, and $\mathcal E$ is a set of edges. Specifically, we examine a scenario that satisfies two following assumptions:
	\begin{asm}\label{asm:graph}\it
		The graph $\mathcal G$ is undirected and connected.
	\end{asm}
	\begin{asm}\label{asm:identifier}\it
		Each agent $i\in\mathcal N$ has a unique identifier $a_i\in\Z_+$ in the sense that $a_i\neq a_j$ for all $j\in\mathcal N\setminus\{i\}$.
	\end{asm}
	
	As a consequence of Assumption \ref{asm:identifier}, two relations can be obtained. First, due to the uniqueness of the identifiers, there exists only one index $l\in\mathcal N$ such that
	\begin{align}
		a_l=a_{\max}\label{eq:a_l-a_max}
	\end{align}
	where $a_{\max}:=\max_{i\in\mathcal N}a_i$ is the largest identifier. Second, there are $N-1$ positive integers smaller than $a_{\max}$, which are  elements in $\{a_i\}_{i\in\mathcal N\setminus \{l\}}$. Thus, we have
	\begin{align}
		a_{\max}\geq N.\label{eq:a_max}
	\end{align}
	
	In the following subsections, we present a distributed algorithm that estimates the network size, characterized by the features (F1) and (F2). We also provide rigorous proofs to demonstrate that all agents can obtain the network size within a finite time.

	\subsection{Proposed Algorithm}
	The proposed algorithm is outlined in Algorithm \ref{algo}. Under this, all agents execute identical dynamic systems using their individual identifiers.  
	\begin{algorithm}
		\caption{Each agent $i$ performs:}\label{algo}
		{\bf Preset:} coupling gain $\gamma>0$, identifier $a_i\in {\mathbb Z}_+$
		
		\vspace{-0.22cm}
		\noindent\rule{\columnwidth}{0.5pt}
		{\bf Initialization:} $z_i(0)\in\R,\quad \mu_i(0)\in\R,\quad x_i(0)\in\R$
		\vspace{0.1cm}
		
		{\bf Dynamics:}
		\begin{itemize}
			\item Estimation for maximum identifier $a_{\max}$:
			\begin{subequations}\label{eq:findMaxID}
				\begin{align}
					\dot z_i &= -z_i + 4a_i^2\max\{a_i-z_i,0\}\notag\\
					&\quad +\gamma\sum_{j\in\mathcal N_i}(z_j-z_i)+\gamma\sum_{j\in\mathcal N_i}(\mu_j-\mu_i)\label{eq:z_i}\\
					\dot \mu_i&=-\gamma\sum_{j\in\mathcal N_i}(z_j-z_i)\label{eq:mu_i}
				\end{align}
			\end{subequations}
			yielding
			\begin{align}
				u_i(t)&:=\lceil z_i(t)\rfloor \label{eq:estimationOutcome}
			\end{align}
			\item Estimation for network size $N$:
			\begin{align}
				\hspace{-0.5cm}\dot x_i =1-I_{a_i}\big(u_i(t)\big)x_i+\big(u_i(t)\big)^3\sum_{j\in\mathcal N_i}(x_j-x_i)\label{eq:x_i}
			\end{align}
			where 
			\begin{align*}
				I_{a_{i}}\big(u_i(t)\big):=\begin{cases}1,\quad &\text{if } u_i(t)=a_i,\\ 0,\quad &\text{otherwise}\end{cases}
			\end{align*}
		\end{itemize}
		{\bf Communicate:} $z_i(t)$, $\mu_i(t)$, $x_i(t)$ \\
		{\bf Output:} $\lceil x_i(t)\rfloor$
	\end{algorithm}
	
	Algorithm \ref{algo} consists of two dynamics: \eqref{eq:findMaxID} and \eqref{eq:x_i}. The first one, represented by \eqref{eq:findMaxID}, aims to find the largest identifier $a_{\max}$. The second one, denoted by \eqref{eq:x_i}, estimates the network size $N$ using the estimation outcome $u_i(t)$ from \eqref{eq:findMaxID}, as defined in \eqref{eq:estimationOutcome}. The purpose of this setup is to transform the system \eqref{eq:x_i} into the form \eqref{eq:exampleForBlended} with the individual vector fields  \eqref{eq:blendedDynamics_N} and a coupling gain $k\geq N^3$. Specifically, when all the variables $u_i(t)$ are equal to the value of $a_{\max}$, it follows from \eqref{eq:a_l-a_max} that $I_{a_i}\big(u_i(t)\big)=\delta_{il}$ for all $i\in\mathcal N$. Furthermore, each coupling gain of \eqref{eq:x_i} satisfies $\big(u_i(t)\big)^3=a_{\max}^3\geq N^3$ for all $i\in\mathcal N$, from \eqref{eq:a_max}. These relations yield that the blended dynamics of \eqref{eq:x_i} becomes \eqref{eq:blendedDynamics}, and hence, the blended dynamics theory ensures that each state $x_i(t)$  approximates the network size $N$.
	
	The form of dynamics \eqref{eq:findMaxID} is derived using the primal-dual gradient algorithm described in \cite{wang2010control} for addressing distributed optimization problems. In particular, we devise a specially tailored optimization problem aimed at estimating the largest identifier with exponential convergence rate, as outlined below:
	\begin{subequations}\label{eq:proposedOptimizationProblem}
		\begin{align}
			\min_{z_1,\cdots, z_N\in\R}\quad &\frac{1}{2}\sum_{i=1}^N \Big( z_i^2+\rho_iP(a_i-z_i)\Big)\label{eq:optimizationCost}\\
			\textnormal{subject to}\quad &\sum_{j\in\mathcal N_i}(z_j-z_i)=0,\quad \forall  i\in\mathcal N\label{eq:OptimizationConstrant}
		\end{align}
	\end{subequations}
	where $\rho_i:=4a_i^2>0$ is the penalty weight and $P:\R\rightarrow\R$ is the penalty function designed by 
	\begin{align*}
		P(p):=\begin{cases}p^2,\quad &\textnormal{if }p\geq 0,\\ 0,\quad &\textnormal{otherwise.}\end{cases}
	\end{align*}
	
	To bring the optimal solutions $z_i^*$ of \eqref{eq:proposedOptimizationProblem} close to the largest identifier $a_{\max}$, the cost function \eqref{eq:optimizationCost} incorporates two terms. The first term $z_i^2$ minimizes the absolute value of $z_i^*$, while the second term $\rho_i P(a_i-z_i)$ penalizes cases where $z_i^*$ is smaller than $a_i$. Moreover, the constraint \eqref{eq:OptimizationConstrant} ensures consensus among all the optimal solutions $z_i^*$.
	
	Note that the dynamics \eqref{eq:findMaxID} exhibits exponential convergence, thanks to the strong convexity of the cost function \eqref{eq:optimizationCost}. This enables us to determine the time required for all states $z_i(t)$ to approach within a range of 0.5 from $a_{\max}$. Once this condition is met, the agents can obtain the integer value $a_{\max}$ by rounding off $z_i(t)$.
	
	\begin{rem}\it
		Several algorithms have been proposed for distributed estimation of the largest identifier. However, none of them successfully achieve both features (F1) and (F2). For example, the approaches in \cite{tahbaz2006one,muniraju2019analysis} utilize the maximum-consensus technique but fail to fulfill feature (F2). Specifically, when the agent with the largest identifier leaves the network, the correct estimation cannot be made without a re-initialization process. On the other hand, algorithms in the papers \cite{deplano2021dynamic,venkategowda2020privacy} do not rely on an initialization process, but their designs require global information such as the network diameter or the network size, respectively, which prevents them from achieving feature (F1).		
	\end{rem}

	\subsection{Analysis and Main Result}\label{sec:mainResult}
	The objective of this subsection is to show that all the agent obtain the exact value of network size $N$ using Algorithm \ref{algo}. To show this, we initially analyze the convergence of the system \eqref{eq:findMaxID}. Specifically, it will be seen that all outcomes $u_i(t)$ of \eqref{eq:findMaxID} converge to the largest identifier $a_{\max}$ within a finite time. Building upon this result, we then show how the dynamics \eqref{eq:x_i} facilitates the network size estimation.
	
	Let us define $z(t) := \text{col}_{i=1}^N \big(z_i(t)\big)\in\R^N$, $\mu(t) := \text{col}_{i=1}^N \big(\mu_i(t)\big)\in\R^N$ and consider the following coordinate changes
	\begin{align}\label{eq:def_coordinate}
		\begin{bmatrix} \bar z(t)\\ \widetilde z(t)\end{bmatrix} :=\begin{bmatrix} \frac{1_N^T}{N}\\ R^T\end{bmatrix}z(t),\quad \begin{bmatrix} \bar \mu(t)\\ \widetilde \mu(t)\end{bmatrix} :=\begin{bmatrix} \frac{1_N^T}{N}\\ R^T\end{bmatrix}\mu(t).
	\end{align} 
	Then, the dynamics  \eqref{eq:findMaxID} can be rewritten as
	\begin{subequations}\label{eq:zbar_ztilde_mubar_mutilde} 
		\begin{align}
			\dot {\bar z} &= -\bar z + \frac{1_N^T}{N}H\Big(1_N\bar z+R\widetilde z\Big)\label{eq:dotbarz}\\
			\dot {\widetilde z} &=-\widetilde z+ R^TH\Big(1_N\bar z+R\widetilde z\Big)-\gamma\Lambda \widetilde z -\gamma\Lambda\widetilde \mu\label{eq:dottildez} \\
			\dot {\widetilde \mu} &= \gamma\Lambda\widetilde z \label{eq:dottildemu}
		\end{align}
	\end{subequations}
	where we use the relations $z(t)=1_N\bar z(t) +R\widetilde z(t)$, $\mu(t)=1_N\bar \mu(t)+R\widetilde \mu(t)$, and the function $H:\R^N\rightarrow \R^N$ is defined as
	\begin{align}
		H(z):=\text{col}_{i=1}^N \big(4a_i^2\max\{a_i-z_i,0\}\big).\label{eq:H}
	\end{align}
	We neglect the state $\bar \mu(t)$ since $\dot {\bar \mu} = 0$, and so, it remains constant over time, i.e,  $\bar \mu(t) = 1_N^T\mu(0)/N$ for all $t\geq 0$. 
	
	Now, we establish exponential convergence of system \eqref{eq:zbar_ztilde_mubar_mutilde}, whose proof can be found in Appendix \ref{apdx:proofOfThm1}.	
	\begin{thm}\label{thm:expOfZ}\it
		Define an error vector $\eta(t)\in\R^{2N-1}$ as follows:
		\begin{align}\label{eq:eta}
			\eta(t) := \begin{bmatrix} z(t)-1_N\bar z^*\\ \widetilde \mu(t)-\widetilde \mu^*\end{bmatrix}
		\end{align}
		where 
		\begin{subequations}\label{eq:equilibriumPoint}
			\begin{align}
				\bar z^*&=\frac{4a_{\max}^3}{N+4a_{\max}^2}~\in \Big(a_{\max}-\frac14,a_{\max}\Big)\label{eq:zbar_star}\\
				\widetilde \mu^*&=\frac{1}{\gamma}\Lambda^{-1}R^TH(1_N\bar z^*)~ \in\R^{N-1}.
			\end{align}
		\end{subequations}
		Under Assumptions \ref{asm:graph} and \ref{asm:identifier},  the solutions of \eqref{eq:findMaxID}, with arbitrary initial conditions, satisfy
		\begin{align}
			|z_i(t)-\bar z^*|< \sqrt 2 |\eta(0)| e^{-\beta t},\quad \forall t\geq 0,~ i\in\mathcal N\label{thm1:result1}
		\end{align}
		where 
		\begin{align}\label{eq:beta}
			\beta :=\frac{\gamma\lambda_2}{4a_{\max}^2+\frac{2(4a_{\max}^2+1)^2}{\gamma\lambda_2}+\gamma\lambda_2+6}.
		\end{align}		
		Moreover, there exists a constant $T_1>0$ such that the variable $u_i(t)$ defined in \eqref{eq:estimationOutcome} satisfies
		\begin{align}\label{thm1:result2}
			\hspace{-0.3cm} u_i(t)=a_{\max},~\forall t\geq  T_1:=\frac{1}{\beta}\ln\big(4\sqrt 2|\eta(0)|\big),~i\in\mathcal N.
		\end{align}
	\end{thm}\vspace{0.2cm}
	
	Next, let us analyze the convergence of the dynamics \eqref{eq:x_i} based on the result of Theorem \ref{thm:expOfZ}. Note that the dynamics \eqref{eq:x_i} can be seen as a time-varying linear system, where the parameters $I_{a_i}\big(u_i(t)\big)$ and $\big(u_i(t)\big)^3$ converge to constants after a sufficient time elapsed. Specifically, it follows from the relations \eqref{eq:a_l-a_max} and \eqref{thm1:result2} that, for all $i\in\mathcal N$ and $t\geq T_1$,
	\begin{subequations}\label{eq:resultofMaxEstimation}
		\begin{align}
			\text{(i)}\quad &I_{a_i}\big(u_i(t)\big)=\delta_{il}\label{eq:resultofMaxEstimation1}\\
			\text{(ii)}\quad &\big(u_i(t)\big)^3=a_{\max}^3.\label{eq:resultofMaxEstimation2}
		\end{align}	
	\end{subequations}
		
	With $x(t):=\text{col}_{i=1}^N(x_i(t))\in\R^N$, the stacked form of \eqref{eq:x_i} can be written as the following linear system:
	\begin{align}
		\dot x = 1_N-\Big(J+a_{\max}^3\mathcal L\Big)x,\quad \forall t\geq T_1\label{eq:stackedDynamics}
	\end{align}	
	where $J\in\R^{N\times N}$ is the matrix whose $l$th diagonal element is $1$ and the rest elements are $0$.
	
	In the next lemma, we establish several relations for the system \eqref{eq:stackedDynamics}, whose proof can be found in Appendix \ref{apdx:lemmas}.
	
	\begin{lem}\label{lem:example}\it
		Under Assumptions \ref{asm:graph} and \ref{asm:identifier}, the matrix $(J+a_{\max}^3\mathcal L)$ is positive definite and satisfies
		\begin{align}
			\lambda_{\min}(J+a_{\max}^3\mathcal L)\geq \frac{1}{4N}\label{eq:minLamOfJL}.
		\end{align}
		Furthermore, the vector $x^*:=(J+a_{\max}^3\mathcal L)^{-1}1_N\in\R^N$ serves as an equilibrium point for the system \eqref{eq:stackedDynamics}, and the $i$th element $x_i^*\in\R$ of $x^*$ satisfies 
		\begin{align}
			\Big|x_i^*-N\Big|< \frac{\sqrt 2}{4},\quad \forall i\in \mathcal N.\label{eq:uboundOfwi-N}
		\end{align}
	\end{lem}\vspace{0.2cm}
		
	Putting all the findings together, we now present the main result of this paper, as follows:
	
	\begin{thm}\label{thm:MainResult}\it
		Suppose that Assumptions \ref{asm:graph} and \ref{asm:identifier} hold. For arbitrary initial conditions $z_i(0)\in\R$, $\mu_i(0)\in\R$, and $x_i(0)\in\R$, there exists a constant $c>0$ such that the solutions of \eqref{eq:x_i} satisfy
		\begin{align}
			|x_i(t)-N|<  ce^{-\frac{1}{4N}t}+\frac{\sqrt 2}{4},~~ \forall t\geq 0,~i\in\mathcal N.\label{thm2:result1}
		\end{align}
		Moreover, it holds that
		\begin{align}
			\hspace{-0.2cm}\lceil x_i(t)\rfloor=N,\quad \forall t\geq T_2:=4N\ln\bigg(\frac{4 c}{2-\sqrt 2}\bigg),~i\in\mathcal N.\label{thm2:result2}
		\end{align}
	\end{thm}\vspace{0.2cm}
	\begin{proof}
		Define an error variable $e_x(t):=x(t)-x^*$ whose time-derivative is given by
		\begin{align}
			\dot e_x&=-(J+a_{\max}^3\mathcal L)e_x+\Delta_x,\quad \forall t\geq 0\label{eq:e_x}
		\end{align}
		where 
		\begin{align}
			\Delta_x(t)&:=\Big[J-\text{diag}_{i=1}^N\Big(I_{a_i}\big(u_i(t) \big)\Big)\Big](e_x(t)+x^*)\notag\\
			&+\Big[\text{diag}_{i=1}^N\Big(a_{\max}^3-\big(u_i(t)\big)^3\Big)\Big]\mathcal L (e_x(t)+x^*).\label{eq:def_Delta_x}
		\end{align}
		
		Meanwhile, it follows from  \eqref{eq:uboundOfwi-N} of Lemma \ref{lem:example} that 
		\begin{align}
			|x_i(t)-N|&\leq |x_i(t)-x_i^*|+|x_i^*-N|\notag\\
			&< |e_x(t)|+\frac{\sqrt 2}{4},\quad \forall i\in\mathcal N.\label{eq:uboundOfxi-N}
		\end{align}
		Thus, the inequality \eqref{thm2:result1} can be proved by showing the exponential stability of \eqref{eq:e_x}. 
		
		As a first step to obtain the exponential convergence of $e_x(t)$, we claim that $\Delta_x(t)$ satisfies the followings:
		\begin{align}
			\text{(i)}~&\text{there exist positive constants $c_1$ and $c_2$ such that}\notag\\
			&\qquad |\Delta_x(t)|\leq c_1|e_x(t)|+c_2,\quad \forall t\geq 0,\label{eq:uboundOfDeltax}\\
			\text{(ii)}~&\Delta_x(t)=0\text{~holds},\quad \forall t\geq T_1 \label{eq:deltax0}
		\end{align}
		where $T_1$ is defined in \eqref{thm1:result2}.
		
		Note that \eqref{eq:deltax0} is the direct result of \eqref{eq:resultofMaxEstimation}. To show the claim \eqref{eq:uboundOfDeltax}, we derive one relation through \eqref{eq:uboundOfL} and \eqref{eq:def_Delta_x}, as follows:
		\begin{align}
			|\Delta_x&(t)|\!\leq\! \Big[1+N\max_{i\in\mathcal N}\big|a_{\max}^3-\big(u_i(t)\big)^3\big|\,\Big]\,|e_x(t)+x^*|\notag\\
			&\!\leq\! \Big[\,1+Na_{\max}^3+N\max_{i\in\mathcal N}\big|u_i(t)\big|^3\,\Big]\,|e_x(t)+x^*|.\label{eq:uboundOfDeltax1}
		\end{align} 
		Meanwhile, it holds from \eqref{eq:estimationOutcome}, \eqref{eq:zbar_star}, and \eqref{thm1:result1} that, for all $i\in\mathcal N$,
		\begin{align}
			|u_i(t)|&\leq |z_i(t)|+\frac12\leq |z_i(t)-\bar z^*|+|\bar z^*|+\frac12\notag\\
			&\leq \sqrt 2|\eta(0)|+a_{\max}+\frac12,\quad \forall t\geq 0.\label{eq:interClaim}
		\end{align}
		By applying \eqref{eq:interClaim} into \eqref{eq:uboundOfDeltax1}, we obtain
		\begin{align}
			&|\Delta_x(t)|\notag\\
			&\!\leq\! \underbrace{\bigg[1\!+\!Na_{\max}^3\!+\!N\Big(\sqrt 2|\eta(0)|\!+\!a_{\max}\!+\!\frac12\Big)^3\bigg] }_{=:c_1}\!\Big(|e_x(t)|\!+\!|x^*|\Big).\label{eq:uboundOfDeltax2}
		\end{align}
		With the constants $c_1$ defined in \eqref{eq:uboundOfDeltax2} and $c_2:=c_1|x^*|$, we complete the proof of the claim \eqref{eq:uboundOfDeltax}.
		
		We now consider a Lyapunov candidate $W:=\frac12 |e_x|^2$, whose time derivative along \eqref{eq:e_x} satisfies 
		\begin{align}
			\dot W &= -e_x^T(J+a_{\max}^3\mathcal L)e_x+e_x^T\Delta_x\notag\\
			&\leq -\frac{1}{4N}|e_x|^2+|e_x||\Delta_x|\label{eq:uBoundOfW}
		\end{align}
		where we use a relation $\lambda_{\min}(J+a_{\max}^3\mathcal L)\!\geq\! 1/4N$ that is obtained from \eqref{eq:minLamOfJL} of Lemma \ref{lem:example}. Meanwhile, the relation \eqref{eq:uboundOfDeltax} yields
		\begin{align*}
			|e_x(t)||\Delta_x(t)|\!&\leq \!c_1|e_x(t)|^2\!+\!c_2|e_x(t)|\!\leq \!(c_1+c_2)|e_x(t)|^2\!+\!c_2
		\end{align*}
		where we use the inequality $|e_x(t)|\leq |e_x(t)|^2+1$.	By applying this into \eqref{eq:uBoundOfW}, we obtain
		\begin{align*}
			\dot W\leq \underbrace{\Big(-\frac{1}{2N}+2c_1+2c_2\Big)}_{=:c_3}W+c_2
		\end{align*}
		which yields\footnote{For nonzero constants $a,b\in\R$, the solution of $\dot h = ah+b$ is given by $h(t)=(h(0)+b/a)e^{at}-b/a$.}
		\begin{align}
			W(t)\leq \Big(W(0)+\frac{c_2}{c_3}\Big)e^{c_3t}-\frac{c_2}{c_3},\quad \forall t\geq 0.\label{eq:uboundOfW}
		\end{align}
		Note that $c_3>0$. Thus, it holds that
		\begin{align}
			W(t)\!\leq c_4:=\Big(W(0)\!+\!\frac{c_2}{c_3}\Big)e^{c_3T_1}\!-\!\frac{c_2}{c_3},~~ \forall t\in[0,T_1].\label{eq:uboundOfW2}
		\end{align}
		
		Meanwhile, by applying \eqref{eq:deltax0} into \eqref{eq:uBoundOfW}, it is derived that
		\begin{align}
			\dot W\leq -\frac{1}{4N}|e_x|^2= -\frac{1}{2N}W,\quad \forall t\geq T_1.\label{eq:uboundOfdotW}
		\end{align}
		Therefore, by combining \eqref{eq:uboundOfW2} and \eqref{eq:uboundOfdotW}, we have
		\begin{align}
			&W(t)\leq c_4e^{-\frac{1}{2N}(t-T_1)},\quad \forall t\geq 0.\label{eq:uboundOfW04}
		\end{align}		
		Now, let us define a constant $c:=\sqrt {2c_4}e^{\frac{T_1}{4N}}$.	Then, by applying \eqref{eq:uboundOfW04} and the relation $|e_x(t)|= \sqrt {2W(t)}$ into the inequality \eqref{eq:uboundOfxi-N}, we have
		\begin{align}
			|x_i(t)-N|&< \sqrt {2W(t)}+\frac{\sqrt 2}{4}\notag\\
			&\leq ce^{-\frac{1}{4N}t}+\frac{\sqrt 2}{4},\quad \forall t\geq 0\label{eq:MainResult}
		\end{align}
		which is equivalent to \eqref{thm2:result1}.
		
		Now, let us show \eqref{thm2:result2}. Define $T_2:=4N\ln\Big(\frac{4 c}{2-\sqrt 2}\Big)$ so that $ce^{-\frac{T_2}{4N}}=\frac{2-\sqrt 2}{4}$. Then, the inequality \eqref{eq:MainResult} becomes
		\begin{align*}
			|x_i(t)-N|<\frac12,\quad \forall t\geq  T_2,~i\in\mathcal N
		\end{align*}
		which proves the relation \eqref{thm2:result2}.
	\end{proof}
	
	\section{Simulation}\label{sec:sim}
	\begin{figure}[t]
		\centering
		\includegraphics[width=0.45\textwidth]{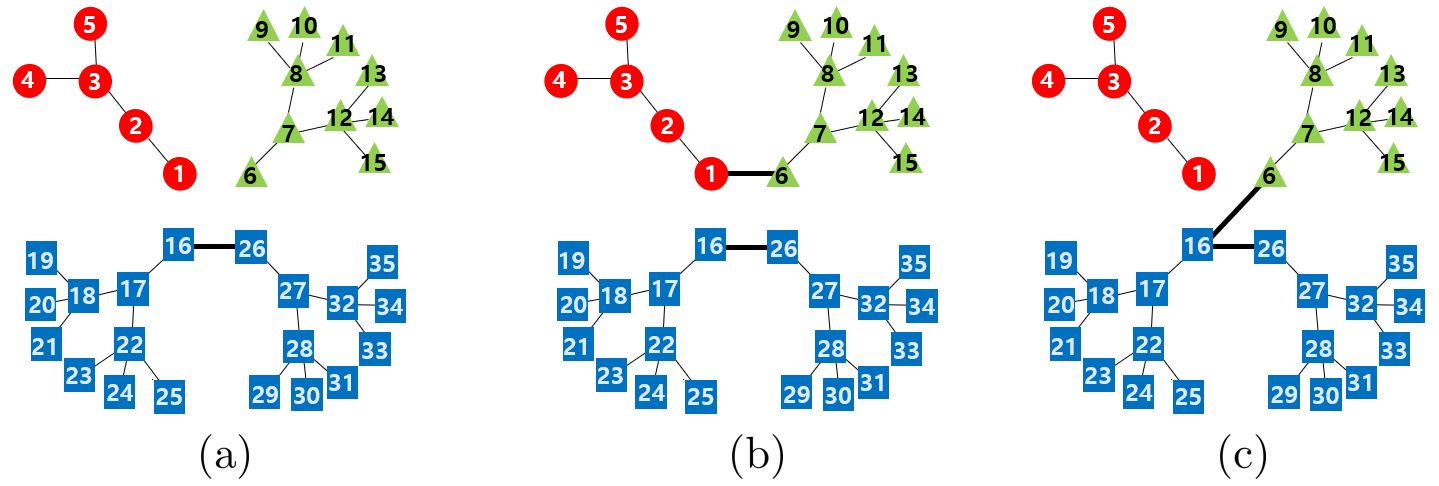}
		\caption{The network is initially represented by the graph (a) and remains for $t\in[0,200)$. When edge $(1,6)$ is added at $t=200$, the network is given by the graph (b). Subsequently, upon the removal of edge $(1,6)$ and the addition of new edge $(6,16)$, the network changes to the graph (c).}
		\label{fig:sim}
	\end{figure}
	Let us consider a communication network of $35$ agents, where each agent $i$ has a  unique identifier $a_i=i$. The network experiences a series of changes, detailed as follows.
	\begin{enumerate}
		\item[1)] Initially, the network consists of three independent connected components\footnote{An independent connected component of a undirected graph $\mathcal G=\{\mathcal N,\mathcal E\}$ is the maximal subgraph $\bar {\mathcal G}:=\{\bar {\mathcal N}, \bar {\mathcal E}\}$ that is connected and such that there is no edge $(i,j)$ in $\mathcal E$ satisfying $i\in\bar {\mathcal N}$ and $j\in\mathcal N\setminus \bar {\mathcal N}$.}: 5 red circle agents, 10 green triangle agents, and 20 blue rectangle agents.
		\item[2)] At $t=200s$, the addition of edge $(1,6)$ combines the connected components of the red and green agents.
		\item[3)] At $t=400s$, the edge $(1,6)$ is removed and a new edge $(6,16)$ is added, linking the green and blue agent components while separating the red agent component.
	\end{enumerate}
	The network for each time interval is illustrated in Fig. \ref{fig:sim}.
	
\begin{figure}[t]
		\centering
		\includegraphics[width=0.45\textwidth]{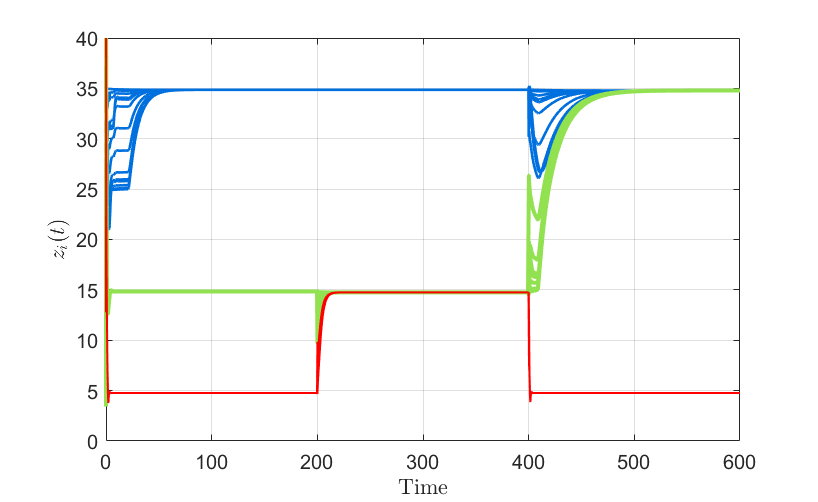}
		\caption{States $z_i(t)$ for largest identifier estimation are represented as dashed lines. They are color-coded: red if $i\in\{1,\cdots,5\}$, green if $i\in\{6,\cdots,15\}$, and blue otherwise.}
		\label{fig:sim_z}
	\end{figure}

\begin{figure}[t]
		\centering
		\includegraphics[width=0.45\textwidth]{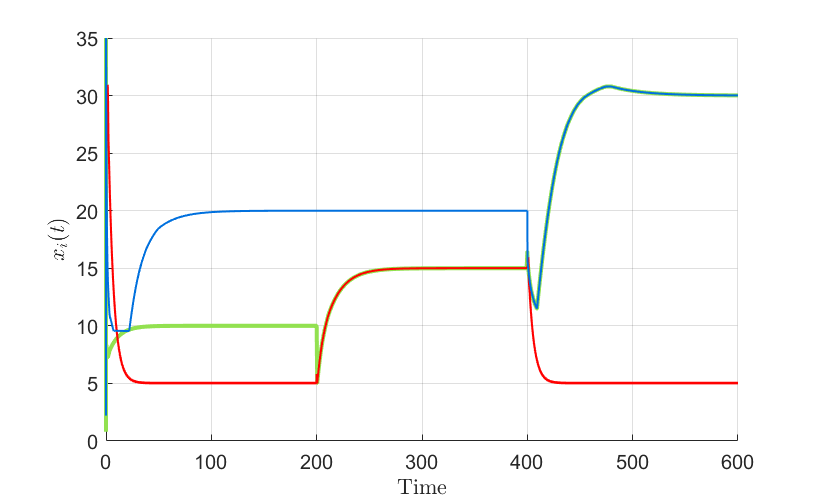}
		\caption{States $x_i(t)$ for network size estimation are illustrated as dashed lines. They are color-coded: red if $i\in\{1,\cdots,5\}$, green if $i\in\{6,\cdots,15\}$, and blue otherwise.}
		\label{fig:sim_x}
	\end{figure}

	All agents in the network execute Algorithm \ref{algo} with the coupling gain $\gamma=10$. Simulation results are depicted in Fig. \ref{fig:sim_z} and Fig. \ref{fig:sim_x}. The results show that each agent obtains the maximum identifier of the agents within the independent connected component it belongs to, which allows to obtain the total number of connected agents.
	
	\section{Conclusion}\label{sec:conclusion}
	This paper has presented a novel distributed estimation scheme for determining the network size of networked multi-agent systems. Inspired by the blended dynamics theory, we propose a distributed algorithm that operates without relying on specific initial conditions. Notably, the incorporation of unique identifiers for each agent allows to design the algorithm in a fully decentralized manner, which includes leader election and coupling gain determination using only local information. In comparison to existing network size estimation techniques in \cite{shames2012distributed,lee18CDC,tran2022distributed}, our approach provides practical benefits by enabling the ``plug-and-play'' operation, wherein any agents can effortlessly join or leave the network without the need for cumbersome re-initialization procedures.
	
	\bibliographystyle{IEEEtran}
	\bibliography{cdsl}
	
	\appendix

	\subsection{Proof of Theorem \ref{thm:expOfZ}}\label{apdx:proofOfThm1}
	
	We first present a technical lemma to prove Theorem \ref{thm:expOfZ}.
	\begin{lem}\label{lem:H}\it
		Consider the function $H:\R^N\rightarrow \R^N$ defined in \eqref{eq:H}. There exists a matrix-valued function $A:\R^N\times \R^{N}\rightarrow \R^{N\times N}$ such that, for all $w\in\R^N$ and $w^*\in\R^N$, the followings hold:
		\begin{align}
			\textnormal{(i)}~ &H(w)-H(w^*)=-A(w,w^*)\cdot(w-w^*)\label{eq:H-Hstar}\\
			\textnormal{(ii)}~ &A(w,w^*)\succeq 0\label{eq:PSDofA}\\
			\textnormal{(iii)}~ &\|A(w,w^*)\|\leq 4a_{\max}^2.\label{eq:UBofA}
		\end{align}
	\end{lem}\vspace{0.2cm}
	\begin{proof}
		For two scalars $y\in\R$ and $y^*\in\R$, let us define a function $\alpha:\R\times \R \rightarrow \R$ as follows:
		\begin{align*}
			\alpha(y,y^*) = \begin{cases}
				\frac{\max\{y,0\}-\max\{y^*,0\}}{y-y^*},\quad &\text{if } y\neq y^*\\
				0,\quad &\text{otherwise.}
			\end{cases}
		\end{align*}
		Then, for all $y,y^*\in\R$, it holds that  
		\begin{align}
			\alpha(y,y^*)&\in[0,1],\label{eq:ubound_alpha}\\
			\max\{y,0\}-\max\{y^*,0\}&=\alpha(y,y^*) \big(y-y^*\big)\label{eq:max_alpha}.
		\end{align}
		From the definition \eqref{eq:H} and the equality \eqref{eq:max_alpha}, we obtain 
		\begin{align}
			H(&w)-H(w^*)\notag\\
			&=\begin{bmatrix} 4a
				_1^2\Big(\max\{a_1-w_1,0\}-\max\{a_1-w_1^*,0\}\Big)\\ \vdots \\ 4a_N^2\Big(\max\{a_N-w_N,0\}-\max\{a_N-w_N^*,0\}\Big)\end{bmatrix}\notag\\
			&=\begin{bmatrix} -4a_1^2\alpha(a_1-w_1,a_1-w_1^*)(w_1-w_1^*)\\ \vdots \\ -4a_N^2\alpha(a_N-w_N,a_N-w_N^*)(w_N-w_N^*)\end{bmatrix}\label{eq:Hz-Hzstar}
		\end{align}
		where $w_i$ and $w_i^*$ are the $i$th element of $w$ and $w^*$, respectively. With a matrix-valued function $A:\R^N\times\R^N\rightarrow \R^{N\times N}$ defined as
		\begin{align} 			\hspace{-0.2cm}A(w,w^*)\!:=\!\text{diag}_{i=1}^N\!\Big(4a_i^2\alpha(a_i\!-\!w_i,a_i\!-\!w_i^*)\Big)\in\R^{N\times N}\label{eq:matA}
		\end{align}
		the equation \eqref{eq:Hz-Hzstar} can be rewritten as
		\begin{align*}
			H(w)-H(w^*)&=-A(w,w^*)\cdot(w-w^*)
		\end{align*}
		which is equivalent to \eqref{eq:H-Hstar}. Meanwhile, \eqref{eq:PSDofA} holds since every diagonal element of $A(w,w^*)$ is non-negative for all $w,w^*\in\R^N$ due to \eqref{eq:ubound_alpha}. The relation \eqref{eq:UBofA} follows from the combination of \eqref{eq:ubound_alpha} and \eqref{eq:matA}. 
	\end{proof}
	
	Now, we will prove Theorem \ref{thm:expOfZ}. The first step is to find the equilibrium point of \eqref{eq:zbar_ztilde_mubar_mutilde}. We start by finding $\widetilde z^*=0$, which follows immediately from the right-hand side of \eqref{eq:dottildemu} since the matrix $\Lambda$ is positive definite. 
    
    Next, we need to determine $\bar z^*$ such that the right-hand side of \eqref{eq:dotbarz} becomes zero. When $\widetilde z^*=0$, this is equivalent to finding a solution $r^*$ of the equation $g(r^*)=0$, where the function $g:\R\rightarrow\R$ is defined as follows:
	\begin{align*}
		g(r)&:=-r+\frac{1_N^T}{N}H(1_Nr)\\
		&=-r+\frac{1}{N}\sum_{i=1}^N 4a_i^2\max\{a_i-r,0\}.
	\end{align*}
	It is worth noting that the function $g$ is continuous, strictly decreasing, positive for a sufficiently large negative $r$, and negative for a sufficiently large positive $r$. Therefore, the existence and uniqueness of  $r^*$ are established. 

    We claim that the solution $r^*$ of $g(r^*)=0$ is given by
	\begin{align*}
		r^*=\frac{4a_{\max}^3}{N+4a_{\max}^2}&=a_{\max}-\frac{1}{\frac{1}{a_{\max}}+\frac{4a_{\max}}{N}}.
	\end{align*}
    To show this, we note that the relation \eqref{eq:a_max} implies
    \begin{align*}
	   \frac{1}{a_{\max}}+\frac{4a_{\max}}{N}>4.
    \end{align*}
    This inequality leads to $r^*\in(a_{\max}-1/4,a_{\max})$. Combining this with \eqref{eq:a_l-a_max}, we have
    \begin{align*}
	\max\{a_i-r^*,0\}=\begin{cases}\frac{1}{  \frac{1}{a_{\max}}+\frac{4a_{\max}}{N}  },\quad &\text{if }i=l\\ 0,\quad &\text{otherwise.} \end{cases}
\end{align*}
    Consequently, we obtain
    \begin{align*}
	g(r^*)&=-r^*+\frac{1}{N}\sum_{i=1}^N4a_i^2\max\{a_i-r^*,0\}\\
    &=-r^*+\frac{4a_{\max}^2}{N}\Bigg(\frac{1}{  \frac{1}{a_{\max}}+\frac{4a_{\max}}{N}  } \Bigg)\\
	&=-r^*+\frac{4a_{\max}^3}{N+4a_{\max}^2}=0.
\end{align*}
    This completes the claim, and therefore, $\bar z^*$ is given as \eqref{eq:zbar_star}.
    
    Meanwhile, under $\widetilde z^*=0$, $\widetilde \mu^*=\frac{1}{\gamma}\Lambda^{-1}R^TH(1_N\bar z^*)$ ensures that the right-hand side of \eqref{eq:dottildez} becomes zero. From above observations, we conclude that $(\bar z^*,\widetilde z^*, \widetilde \mu^*)$ is the equilibrium point of \eqref{eq:zbar_ztilde_mubar_mutilde}.
	
	The second step is to analyze the stability of the system \eqref{eq:zbar_ztilde_mubar_mutilde}. Consider the following error variables:
	\begin{align}
		\bar e_{z}&:=\sqrt N(\bar z-\bar z^*),~\widetilde e_{z}:= \widetilde z-\widetilde z^*,~\widetilde e_{\mu}:=\widetilde \mu-\widetilde \mu^*\label{def:errors}
	\end{align}
	whose derivatives can be represented as
	\begin{align}
		\begin{split}\label{eq:ezbar_eztilde_emutilde}
			\dot {\bar e}_{z}&=-\bar e_{z}-\frac{1_N^T}{\sqrt N}\bigg(A(z,z^*)\frac{1_N}{\sqrt N}\bar e_{z}+A(z,z^*)R\widetilde e_{z}\bigg)\\
			\dot {\widetilde e}_{z}&=-\widetilde e_{z}-R^T\bigg(A(z,z^*)\frac{1_N}{\sqrt N}\bar e_{ z}+A(z,z^*)R\widetilde e_{z}\bigg)\\
			&\quad -\gamma\Lambda \widetilde e_{z}-\gamma\Lambda \widetilde e_{\mu}\\
			\dot {\widetilde e}_{\mu}&=\gamma\Lambda \widetilde e_{z}
		\end{split}
	\end{align}
	where $A:\R^N\times \R^N\rightarrow \R^{N\times N}$ is the matrix-valued function obtained from Lemma \ref{lem:H}, and $z^*:=1_N\bar z^*+R\widetilde z^*$. Meanwhile, by using \eqref{eq:aboutLaplacian}, it follows from the definitions \eqref{eq:eta} and \eqref{def:errors} that
	\begin{align}
		\hspace{-0.25cm}|z_i(t)-\bar z^*|^2\leq |\eta(t)|^2=|\bar e_{ z}(t)|^2+|\widetilde e_{ z}(t)|^2+|\widetilde e_{ \mu}(t)|^2.\label{eq:uboundOfez}
	\end{align}
	Therefore, the inequality \eqref{thm1:result1} can be proved by showing the exponential stability of the system \eqref{eq:ezbar_eztilde_emutilde}. 
	
	To this end, we define a Lyapunov candidate $V$ as
	\begin{align*}
		V:=\frac{\phi}{2}|\bar e_{z}|^2 + \frac{\phi}{2} |\widetilde e_{z}|^2 +\frac{(\phi+1)}{2}|\widetilde e_{\mu}|^2 +\frac12|\widetilde e_{z}+\widetilde e_{\mu}|^2
	\end{align*}
	where $\phi>0$ will be chosen later. Here we note that
	\begin{align}
		\frac{\phi}{2}\big(|\bar e_{z}|^2+|\widetilde e_{z}|^2+&|\widetilde e_{ \mu}|^2\big)\leq V \label{eq:LBofV}\\
		&\leq \frac{\phi+3}{2}\big(|\bar e_{z}|^2+|\widetilde e_{z}|^2+|\widetilde e_{\mu}|^2\big).\label{eq:UBofV}
	\end{align}
	Then, the derivative of $V$ along \eqref{eq:ezbar_eztilde_emutilde}  is given by
	\begin{align}
		\dot V &={\color{black}-\phi |\bar e_{z}|^2-(\phi+1)|\widetilde e_z|^2-\phi \gamma \widetilde e_{z}^T\Lambda \widetilde e_{ z}-\gamma \widetilde e_{\mu}^T\Lambda \widetilde e_{\mu}}\notag\\
		&\quad -\phi \Big(\frac{1_N}{\sqrt N}\bar e_{z}+R\widetilde e_{z}\Big)^TA(z,z^*)\Big(\frac{1_N}{\sqrt N}\bar e_{z}+R\widetilde  e_{z}\Big)\notag\\
		&\quad {\color{black}-(R\widetilde e_{z})^T A(z,z^*)R\widetilde e_{z}}\notag\\
		&\quad - \widetilde e_{z}^TR^TA(z,z^*)\frac{1_N}{\sqrt N}\bar e_{z}-\widetilde e_{\mu}^TR^TA(z,z^*)\frac{1_N}{\sqrt N}\bar  e_{z}\notag\\
		&\quad -\widetilde e_{\mu}^T\widetilde e_{z}-\widetilde e_{\mu}^TR^TA(z,z^*)R\widetilde e_{z}\notag\\
		&\leq {\color{black}-\phi |\bar e_{z}|^2-\big(\phi +1 + \phi\gamma\lambda_2\big)|\widetilde e_{z}|^2-\gamma\lambda_2 |\widetilde e_{\mu}|^2}\notag\\
		&\quad +4a_{\max}^2|\widetilde e_{z}||\bar e_{z}|+4a_{\max}^2|\widetilde e_{\mu}||\bar e_{z}|\notag\\
		&\quad +(4a_{\max}^2+1)|\widetilde e_{\mu}||\widetilde e_{z}|\label{eq:UBofDV}\\
		&\leq -\bigg(\phi-2a_{\max}^2-\frac{16a_{\max}^4}{\gamma\lambda_2}\bigg)|\bar e_{z}|^2-\frac{\gamma\lambda_2}{2}|\widetilde e_{\mu}|^2\notag\\
		&\hspace{-0.3cm}-\bigg(\phi(1+\gamma\lambda_2)+1-2a_{\max}^2-\frac{(4a_{\max}^2+1)^2}{\gamma\lambda_2}\bigg)|\widetilde e_{z}|^2\label{eq:UBofDV2}
	\end{align}
	where the inequality \eqref{eq:UBofDV} follows from the relations \eqref{eq:PSDofA} and \eqref{eq:UBofA} of Lemma \ref{lem:H}, and the last inequality \eqref{eq:UBofDV2} is obtained from the followings:
	\begin{align*}
		|\widetilde e_{z}||\bar e_{z}|&\leq \frac12 |\widetilde e_{z}|^2+\frac12 |\bar e_{z}|^2\\
		|\widetilde e_{\mu}||\bar e_{z}|&\leq  \frac{\gamma\lambda_2}{16a_{\max}^2}|\widetilde e_{\mu}|^2+\frac{4a_{\max}^2}{\gamma\lambda_2}|\bar e_{z}|^2\\
		|\widetilde  e_{\mu}||\widetilde e_{z}|&\leq \frac{\gamma\lambda_2}{4(4a_{\max}^2+1)}|\widetilde e_{\mu}|^2+\frac{4a_{\max}^2+1}{\gamma\lambda_2}|\widetilde e_{z}|^2.
	\end{align*}
	Therefore, if $\phi$ satisfies 
	\begin{align*}
		\phi\!>\!\max\Big\{ 2a_{\max}^2\!+\!\frac{16a_{\max}^4}{\gamma\lambda_2}, \frac{1}{1\!+\!\gamma\lambda_2}\!\Big[2a_{\max}^2\!+\!\frac{(4a_{\max}^2\!+\!1)^2}{\gamma\lambda_2}\!\Big]\! \Big\}
	\end{align*}
	then $\dot V$ becomes negative definite. Let us choose $\phi$ as
	\begin{align}
		&\phi:=2a_{\max}^2+\frac{(4a_{\max}^2+1)^2}{\gamma\lambda_2}+\frac{\gamma\lambda_2}{2}\notag\\
		&\geq \! 2a_{\max}^2\!+\!2\sqrt{\frac{(4a_{\max}^2+1)^2}{2}}\!=\! (2\!+\!4\sqrt 2)a_{\max}^2\!+\!\sqrt 2.\label{eq:phi}
	\end{align}
	From this and \eqref{eq:UBofV}, the inequality \eqref{eq:UBofDV2} becomes
	\begin{align}
		\dot V&\leq -\frac{\gamma\lambda_2}{2}\big(|\bar e_{z}|^2+|\widetilde e_{z}|^2+|\widetilde e_{\mu}|^2\big)\leq  -\frac{\gamma\lambda_2}{\phi+3}V. \label{eq:UBofDV3}
	\end{align}
	Thus, by using \eqref{eq:uboundOfez}, \eqref{eq:LBofV}, \eqref{eq:UBofV}, and \eqref{eq:UBofDV3}, we obtain an upper bound of $|\eta(t)|^2$, given by
	\begin{align}
		\hspace{-0.4cm}|\eta(t)|^2\!\leq\!  \frac{2}{\phi}V(t)\!\leq\! \frac{2}{\phi}V(0)e^{-\frac{\gamma\lambda_2}{\phi+3}t}\!\leq\! \frac{\phi+3}{\phi}|\eta(0)|^2e^{-\frac{\gamma\lambda_2}{\phi+3}t}.\label{eq:UBofE2}
	\end{align}
	Under the condition \eqref{eq:phi}, we obtain $(\phi+3)/\phi<2$. Therefore,  by combining \eqref{eq:uboundOfez} and \eqref{eq:UBofE2}, we finally obtain the inequality \eqref{thm1:result1}. 
	
	Now, we prove the equality \eqref{thm1:result2}. From the relations \eqref{eq:zbar_star} and \eqref{thm1:result1}, we have
	\begin{align}
		|z_i(t)-a_{\max}|&\leq |z_i(t)-\bar z^*|+|\bar z^*-a_{\max}|\notag\\
		&< \sqrt 2 |\eta(0)|e^{-\beta t}+\frac14,\quad \forall i\in\mathcal N.\label{eq:cor1}
	\end{align}
	Let us define $T_1:=\ln\big(4\sqrt 2|\eta(0)|\big)/\beta$ so that $\sqrt 2 |\eta(0)|e^{-\beta T_1}=1/4$ holds. Therefore, for all $i\in\mathcal N$, \eqref{eq:cor1} becomes
	\begin{align*}
		|z_i(t)-a_{\max}|<\frac12,\quad \forall t\geq T_1
	\end{align*}
	which is equivalent to \eqref{thm1:result2}.

	\subsection{Proof of Lemma \ref{lem:example}}\label{apdx:lemmas}	
	Note that, it is a direct consequence of \cite[Lemma 1]{lee18CDC} that the matrix $(J+a_{\max}^3\mathcal L)$ is positive definite. Furthermore, the inequality \eqref{eq:minLamOfJL} also follows by \cite[Lemma 1]{lee18CDC} since $a_{\max}^3\geq N^3>N/\lambda_2$ from \eqref{eq:lboundOflambda2} and \eqref{eq:a_max}. 
	
	Now, we focus on the system \eqref{eq:stackedDynamics}. It is evident from the right-hand side of \eqref{eq:stackedDynamics} that the vector $x^*$ serves as an equilibrium point for this system. In order to show \eqref{eq:uboundOfwi-N}, we consider a relation $1_N^T(J+a_{\max}^3\mathcal L)=1_N^TJ=e_l^T$, where $e_l\in\R^N$ is the standard unit vector whose $l$th element is 1 and the rest are 0. Thus, the $l$th row of $(J+a_{\max}^3L)^{-1}$ is $1_N^T$, and this implies that
	\begin{align}
		x_l^*=1_N^T1_N = N. \label{eq:x1_star}
	\end{align}
	
	Meanwhile, let us define $\bar x^*:=1_N^Tx^*/N$ and $\widetilde x^*:=R^Tx^*$. We then obtain that $x^*=1_N\bar x^*+R\widetilde x^*$. By multiplying $R^T(J+a_{\max}^3\mathcal L)$ into both sides of the equation $(J+a_{\max}^3\mathcal L)^{-1}1_N=x^*$, we obtain
	\begin{align}
		R^T1_N &= R^T(J+a_{\max}^3\mathcal L)x^*\notag\\
		&=R^TJx^*+a_{\max}^3R^T\mathcal L x^*\notag\\
		&=R^TJx^*+a_{\max}^3R^T\mathcal L(1_N\bar x^*+R\widetilde x^*)\notag\\
		&=NR^Te_l+a_{\max}^3\Lambda \widetilde x^*\label{eq:tilde_x_star}
	\end{align}
	where the last inequality \eqref{eq:tilde_x_star} follows from \eqref{eq:aboutLaplacian} and \eqref{eq:x1_star}. Since $R^T1_N=0$, the relation \eqref{eq:tilde_x_star} yields
	\begin{align}\label{eq:ubound_tilde_x_star}
		|\widetilde x^*|=\bigg|-\frac{N}{a_{\max}^3}\Lambda^{-1}R^Te_l\bigg|\leq \frac{N}{a_{\max}^3\lambda_2}< \frac{1}{4}
	\end{align}
	where we use \eqref{eq:lboundOflambda2} and \eqref{eq:a_max}.	 
	
	Meanwhile, we note that the set of all columns in $\big[\frac{1_N}{\sqrt N},~R\big]$ is an orthonormal basis of $R^N$. Thus, from the fact $1_N^T(e_i-e_l)=0$, one can find $\theta_i\in\R^N$ such that $|\theta_i|=\sqrt 2$ and
	\begin{align*}
		R\theta_i=e_i-e_l,\quad \forall i\in\mathcal N\setminus \{l\}.
	\end{align*}
	From this, \eqref{eq:x1_star}, and \eqref{eq:ubound_tilde_x_star}, we finally obtain 	
	\begin{align}
		|x_i^*-N|&=|e_i^Tx^*-e_l^Tx^*|=|\theta_i^TR^Tx^*|\leq\sqrt 2|\widetilde x^*|\notag\\
		&< \frac{\sqrt 2}{4},\quad \forall i\in\mathcal N\setminus \{l\}.\label{eq:uboundOfw-N}
	\end{align}
	The relations \eqref{eq:x1_star} and \eqref{eq:uboundOfw-N} completes the proof of \eqref{eq:uboundOfwi-N}.
\end{document}